\documentclass[twocolumn,prl,showpacs,floatfix]{revtex4}
\usepackage{bm}
\usepackage{array}

\newcommand{\gtapprox}{{\raise.3ex\hbox{$>$\kern-.75em\lower1ex\hbox{$\sim$}}}}
\newcommand{\keff}{k}
\newcommand{\lf}{\ell_f}
\newcommand{\bfx}{{\bf x}}

\usepackage[dvips]{epsfig}

\usepackage{float}

\usepackage{amsmath}
\usepackage{amsfonts}

\begin{document}
\author{B.A. DiDonna}
\affiliation{Institute for Mathematics and its Applications,
University of Minnesota, Minneapolis, MN 55455-0436, USA}
\author{Alex J. Levine}
\affiliation{
Department of Chemistry and Biochemistry,
University of California,
Los Angeles, CA 90095, USA}

\title{Filamin cross-linked semiflexible networks: Fragility under strain}

\begin{abstract}
The semiflexible F-actin network of the cytoskeleton is cross-linked
by a variety of proteins including filamin, which contain Ig-domains
that unfold under applied tension. We examine a simple semiflexible
network model cross-linked by such unfolding linkers that captures
the main mechanical features of F-actin networks cross-linked by
filamin proteins and show that under sufficiently high strain the
network spontaneously self-organizes so that an appreciable fraction
of the filamin cross-linkers are at the threshold of domain
unfolding. We propose an explanation of this organization based on a
mean-field model and suggest a qualitative experimental signature of
this type of network reorganization under applied strain that may be
observable in intracellular microrheology experiments of Crocker
{\em et al.}.
\end{abstract}

\date{\today}


\maketitle

The cytoskeleton of eukaryotic cells is a cross-linked  biopolymer
network\cite{alberts:88,pollard:86,elson:88}. Its principal
constituent is a stiff protein aggregate (F-actin) that is
cross-linked densely on the scale of its own thermal persistence
length. Because of the combination of filament stiffness and dense cross-linking this \emph{semiflexible} polymer gel differs
fundamentally from the better understood \emph{flexible} polymer
gels that are the products of modern synthetic chemistry.

There has been considerable progress in understanding
the complex relationship between the mechanical properties of
semiflexible networks and the mechanical properties of their
constituent filaments~\cite{janmey:90,kroy:96,satcher:96,mackintosh:95,
head:many,wilhelm:03,didonna:06,gardel:04}. Since much of this work addresses highly simplified systems, one may ask how well does the current understanding of semiflexible networks elucidate the rheology of the cytoskeleton, which is a highly heterogenous chemical system.  Cytoskeletal filaments are polydisperse in length and have a greater
distribution of mechanical properties (due to \emph{e.g.} filament bundling) than the model semiflexible network systems studied. Furthermore these filaments are cross-linked by a plethora of highly structured proteins that play an active role in generating internal stresses and in sensing externally imposed stress. One class of cross-linking proteins contain numerous repeat domains, such as titin\cite{labeit:95,reif:97} and
filamin\cite{schwaiger:04,brockwell:05} that unfold reversibly at a critical pulling force.

In order to begin to address the mechanical effect of this chemical heterogeneity in the cytoskeleton,  we investigate
networks cross-linked by filamin--like proteins containing multiple unfolding domains. We find that, above a certain strain threshold,the population of cross-links at
given tension grows rapidly up to the critical unfolding
tension of the domains. Thus, the system appears to
self-consistently adjust its mechanical properties so as to reach a highly fragile state in which a large fraction of its
cross-linkers are poised at the unbinding transition of their
internal domains. The evolution of this high susceptibility state in which the system responds to small stress fluctuations with
anomalously large strain fluctuations may contribute to the large
nonthermal low-frequency intracellular strain fluctuations as
measured by Hoffman {\em et el.}\cite{hoffman:05}. The network may evolve into this high susceptibility state under the action of internal molecular motors ({\em e.g.} myosin -- not considered in our model) so that small fluctuations in motor protein
activity leads to the coordinated unfolding of numerous filamin
cross-linkers and the consequent large-scale cytoskeletal
rearrangement event.

\begin{figure}
\centerline{\includegraphics[width=2.5in]{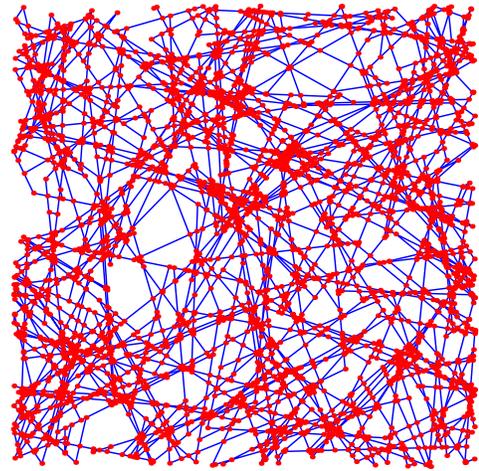}}
\caption{(color online)
Model network showing the F-actin filaments in blue and the
filamin cross-linking agents in red. }
\label{fig:network}
\label{fig:actfilsprings}
\end{figure}

We study via numerical simulation a random, statistically homogeneous, two-dimensional,
isotropic filament network. These networks are
formed in a manner identical to that of Head {\em et
al.}\cite{head:many}. At filament intersections we add a
cross-linker of zero rest length that exerts constraint forces but no constraint torques. A model network constructed by this procedure is shown in Fig.~\ref{fig:network}. The filament sections between
crosslinks are modeled as linear springs with fixed elastic constant {\em per unit length}. The nonlinear behavior of semiflexible networks with freely rotating cross-links has been shown to be dominated by semiflexible filament stretching instead of bending~\cite{onck:05}, so we neglect filament bending. We
anticipate that the results derived here are essentially independent of network dimensionality since network connectivity, not the dimensionality of the space in which the network is embedded, should control the collective mechanical properties of the system.

At forces below the unfolding force, the force extension relation of the filamin cross-linkers is that of a worm-like
chain\cite{marko:95}. When an Ig domain unfolds the contour length of the filamin increases, adding enough length to relax most of the tension at fixed extension. For simplicity, we model the filamin as a simple spring with spring constant $k_f$ and we take the additional contour length generated during any unfolding event $\lf$ to be a constant. The critical unfolding force of a domain is $k_f \lf$. We neglect the the rate-dependence of this unfolding force\cite{evans:97}. Though the physiological filament
cross-linkers have a finite number of unfolding domains ($24$), we will take our sawtooth force extension curve to have an infinite number of branches.

The network is sheared using Lees-Edwards boundary conditions. At
the beginning of each strain step all nodes are moved affinely, then the node positions are relaxed through a conjugate gradient routine to a point of local force equilibrium. Since the cross-linker force extension curve is a sawtooth, there are many possible equilibrium
states of the network. In principal, the multiplicity of equilibrium states
requires us to use strain steps resulting in displacements smaller than the sawtooth length $\lf$ so that equilibrium is achieved at the smallest filamin extension. To reduce computational overhead we
use a two step equilibration procedure that finds a state close to desired one, but allows for large strain steps. In the first
equilibration step, we replace the sawtooth force law for all
cross-linkers by the following force law:
\begin{equation}
{\bf f} =
\begin{cases}
k_f \bfx & \left|\bfx\right| < \lf, \\
k_f \lf  & \left|\bfx\right| \ge \lf.
\end{cases}
\label{eq:flatforce}
\end{equation}
The combined network of linear elastic filaments and constant force cross-links is equilibrated. We then reimpose a sawtooth force law for the cross-linkers and equilibrate the network a second time. As the network relaxes during this final equilibration step, the force on each filamin must be less than $k_f \lf$, so the cross-links will stay on the same sawtooth branch. Since the rest of the network was originally equilibrated at the critical pulling force, the sawtooth
force law could not have reached force equilibrium on any earlier
sawtooth branch assuming all filamin linkers act independently. In practice, collective relaxations of the network push individual cross-links onto different sawtooth branches in this final step. We found, however, that such coordinated relaxation events had a negligible quantitative effect on the data.

We present data for networks composed of monodisperse filaments of length $\ell_R=0.2$ (all lengths are measured in units of the length of the square simulation box) at a filament density such that there are on average $30$ cross-links per filament. For these values we
find negligible system-size effects. The length of the filamin
domains is given by $\lf=1.3 \times 10^{-4}$, or $\lf/\ell_c=0.02$.
This ratio is about ten times smaller than the expected
physiological value~\cite{freykroy:98, furuike:01}. We choose the
smaller value of $\lf/\ell_c$ because it enhances the effects we
were measuring; our initial studies at
physiological values of $\lf/\ell_c$ find qualitatively similar
results, but the onset of non-linear effects occurred at higher
strain values~\footnote{ Non-linear effects may occur at vanishingly small stain in a prestressed network as found in the cytoskeleton. The effects of prestress will be explored in future
work~\cite{Didonna:06}.}. To fix an energy scale we set the
extensional modulus $\mu$ of the filaments to unity. The average
spring constant for a filament segment can then be determined from the mean distance between cross-links: $k_R = 1 /
\left<\ell_c\right> = 150$. The range of cross-linker spring constant values studied here is $10^1 < k_f < 10^4$.

\begin{figure}
\centerline{\includegraphics{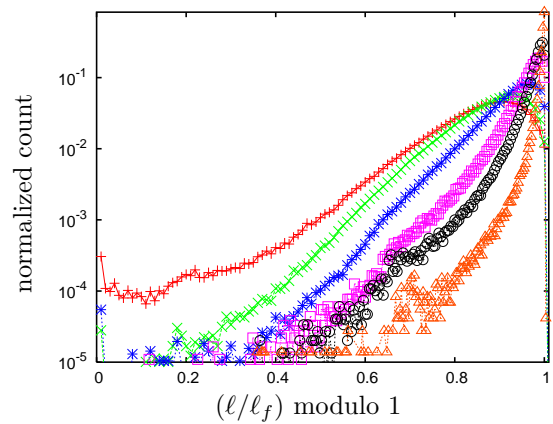}}
\caption{ (color online) Distribution of normalized cross-linker lengths $\ell/\ell_f$ modulo $1$ in equilibrated networks with, 
from shallowest to steepest slopes respectively, $k_f=100$, 
$200$, $600$, $2000$, $4000$, and $10000$.} \label{fig:pileup}
\end{figure}

Figure~\ref{fig:pileup} shows the measured equilibrium distributions
of cross-link lengths, modulo the sawtooth length $\lf$, for a
representative set of strained networks with $\lf=1.3
\times 10^{-4}$ and several values of spring constant $k_f$. For values of  $k_f<10 \times \left<k_R\right>$ the statistical
weight for finding a cross-link extension (modulo $\lf$) is
exponentially enhanced towards length $\lf$ where the domains
unbind. For values $k_f>10 \times \left<k_R\right>$ the statistical weight for finding a cross-link extension (modulo $\lf$) grows faster than exponentially near length $\lf$. 
The enhancement of the number of such filamins poised at the
unbinding transition (critical cross-linkers) is one of principal results of our work. Apart from the noise floor, the shape of the distribution appears to be strain independent. 

To understand the growth of the critical cross-linker population, we consider a mean-field model for the behavior of a single cross-link
in an effective elastic medium representing the rest of the network.
The surrounding effective medium is modeled as a single harmonic
spring with spring constant $k$. Reflecting the network structure,
the cross-linker is connected in series with the effective network
spring. We set the total strain on the two springs in series (by
fixing their total length) so that the cross-link has crossed at
least one branch of the sawtooth function. Upon the further
application of extensional strain, the two springs in series will
both  linearly increase their extension until the filamin spring
with spring constant $k_f$ reaches is maximum force $k_f \ell_f$
where it is poised at the top of its saw-tooth force-extension
curve.

Now consider an infinitesimal increase in the total extension that
drives the unfolding of one more filamin domain. Before the
extension the two springs were in force balance so that $k_f \ell_f
= k x$ where $x$ represents the extension of the medium spring.
After the infinitesimal extension, the system achieves force balance
on the next branch of the saw-tooth filamin force-extension curve so
that extension of the filamin spring is now increased by $\ell_f -
d$ while the extension of the medium spring is decreased to $x -
(\ell_f - d)$. Force balance requires that $d$, the distance between
the current extension of the filamin spring and the edge of the next
saw-tooth, is given by $d(k) = k \ell_f/(k + k_f)$. In other words,
the combined system once equilibrated with the filamin spring at its
maximal force is now equilibrated with that filamin spring on its
next saw-tooth branch at a smaller force. The strain in the
surrounding medium has also decreased due to the extension of one
more filamin domain.

To maintain force balance, the filamin spring cannot relax its
length more than $\ell_f - d$. Upon further extension the filamin
spring will only extend until another domain unbinds. Thus in
steady-state the filamin spring will evenly sample all extensions
(modulo $\ell_f$) between $\ell_f - d(k)$ and $\ell_f$. For a given
value of the spring constant of the medium we expect that the
extensions (modulo $\ell_f$) of the filamin cross-linkers $x_f$ to
be uniformly distributed between the bounds given above so that this
distribution can be written as
\begin{equation}
P\left(x_f,k \right)=\frac{1}{d(k)} \Theta \left(x_f - \left[\lf -
d(k) \right]\right),
\end{equation}
where $\Theta$ is a step-function. Different cross-links in the
network,however,  will not have the same local environments; the
values of $k$ will be sampled from some statistical distribution
$K\left(k \right)$. Integrating over that distribution we write the
probability of finding a given filamin length (modulo $\lf$) $x_f$:
\begin{equation}
P\left(x_f\right)=\int_{k_f \frac{\lf-x_f}{x_f}}^\infty \frac{\keff
+ k_f}{\lf \keff} K\left(\keff\right) d \keff. \label{eq:pxint}
\end{equation}
The step function fixes the lower limit on the $k$-integral.

We examine the distribution of the local
spring constants in the random network and concentrate on the high-$k$ tail of that distribution. One may imagine that the effective spring constant representing the medium can be represented as a
small number of chains of springs. Each chain of springs represents
one path for force propagation through the random network; it is
made up of a large number of statistically independent springs
connected in series. In order to find an extremely large value of
the effective spring constant $k$ it must be that for one of the
force paths {\em all of the constituent spring constants are large},
since the compliance of the springs in series will be dominated by
any single soft spring. We expect the probability of such a rare
event to be Poisson distributed so that, in the
high-$k$ tail, the distribution $K(k)$ takes the form
\begin{equation}
\label{Poisson-tail} K(k) \sim H(k) e^{-k/\bar{k}}
\end{equation}
where $H$ is some regular function characterizing the small-$k$
behavior of the distribution ($H(x) \rightarrow \mbox{const}$ as $x
\rightarrow \infty$) and the constant $\bar{k}$ is undetermined.
Combining Eqs.~\ref{eq:pxint},\ref{Poisson-tail} we find that
$P(x_f)$ takes the form
\begin{multline}
\label{exp-dis} P(x_f)\simeq \frac{\bar{k}}{\lf}
\exp\left(\frac{k_f (x_f - \lf)}{\bar{k} x_f}\right) + \\
\frac{k_f}{\lf} \Gamma
\left(0,\frac{k_f (\lf - x_f)}{\bar{k} x_f}\right)
\end{multline}
as long as $k_f \frac{\lf - x_f}{x_f}$ is large enough that $K(k)$
within the integral in Eq.~\ref{eq:pxint} can be replaced by its
high-$k$ asymptotic form. Eq.\ref{exp-dis} shows the sought after
exponential peak as $x_f \longrightarrow \lf$.

\begin{figure}
\centerline{\includegraphics{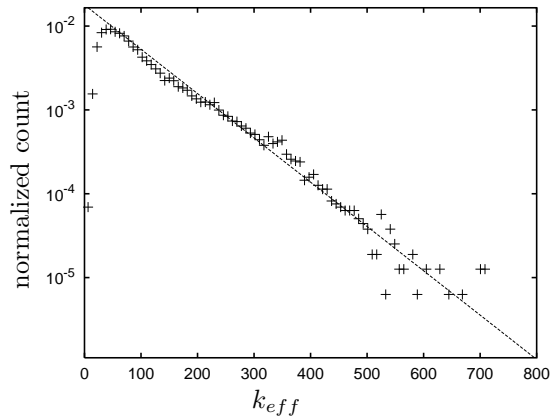}} \caption{Distribution
of local effective spring constants, sampled on small sets of highly
stretched filamin cross-linkers with $k_f=600$. The solid line is a
fit to Eq.~\ref{Poisson-tail} with $\keff/\bar{k}$ set to $7.3$. }
\label{fig:expmoddist}
\end{figure}

Using Eq.~\ref{exp-dis} we may
determine the ratio $k_f/\bar{k}$ using the slope of the numerically
measured distribution of $x_f$ shown in Figure~\ref{fig:pileup}.
From fitting the data for $k_f=600$ we find that $k_f/\bar{k} =7.3$. 
By numerically sampling the local mechanical response in many
realizations of the network, we independently verify the principal
physical insight in the above discussion: for small values of 
$k_f$, $K(k)$ appears to have an
exponential tail in the stiff spring limit.  This data is presented
in Fig.~\ref{fig:expmoddist} for $k_f=600$. 
The fitted line demonstrates that the high $k$ tail of 
this data is consitent with
a value of $\keff/\bar{k}=7.3$. The 
agreement between the two independently determined values of
$\bar{k}$ supports our analysis. In upcoming work, 
we demonstrate that
for higher $k_f$ the distribution function $K(k)$ takes a different form, 
but it is still related to the distribution 
function $P(x_f)$ via Eq.~\ref{eq:pxint}~\cite{Didonna:06}. This latter
equation should apply for general distributions $K(k)$.

We have found that the strained filamin cross-linked network develops into a highly
fragile mechanical state in which a large fraction of the
cross-linking filamins reach a critical state where they are poised at the brink of domain unfolding. The presence of fluctuating internal
stresses in the cytoskeleton produced by molecular motor activity
and/or equilibrium fluctuations can act on this highly fragile state to produce large strain fluctuations due to the correlated failure of numerous critical cross-linkers. Thus, the observation of the formation of this critical state under applied stress may explain a particular feature of the observed low-frequency strain fluctuations as observed by intracellular microrheology.

We have presented a simple, mean-field theory to explain the
evolution of this fragile state under applied strain. There are a
number of extensions of this work that remain to be considered.
Foremost among these is the exploration of the effect of
filamin-type cross-linkers in semiflexible gels where the filaments
each have a finite bending modulus. In addition, the development of a complete model that includes the effect of internally generated random stresses due to the action of molecular motors will be an important step towards the direct calculation of the low-frequency dynamics of this biopolymer gel of fundamental biological importance.


BD and AJL thank J.C. Crocker for providing unpublished data and
for enjoyable discussions. BD also thanks David Morse for
enlightening discussions. AJL was supported in part by
NSF-DMR0354113. BD acknowledges the hospitality of the UCLA
department of Chemistry and the California Nanoscience Institute
where part of this work was done. BD also acknowledges partial
support from NSF-DMR0354113 and the Institute for Mathematics and its Applications with funds provided by the
National Science Foundation.

\end{document}